\documentclass{roajarticle}
\usepackage{graphicx}
\usepackage{mynatbib}
\usepackage{upgreek}
\usepackage{upgreek2}
\usepackage{amsmath}
\usepackage[section]{placeins}
\newcommand{\volume}{\textbf{1}}
\newcommand{\numarul}{1}
\newcommand{\volyear}{2019}
\graphicspath{ {./figs/} }

\title{Jacobi stability analysis of the classical restricted three body problem}
\newcommand{\shorttitle}{Jacobi stability of the classical restricted three body problem}

\author[1]{Cristina  \MakeTextUppercase{Blaga}}
\author[2]{Paul A. \MakeTextUppercase{Blaga}}
\author[3]{Tiberiu \MakeTextUppercase{Harko}}

\affil[1]{Babe\c{s}-Bolyai University, Faculty of Mathematics and Computer Science,
Kog\u{a}lniceanu Street 1, 600410, Cluj-Napoca, Romania,
Email: cpblaga@math.ubbcluj.ro}
\affil[2]{ Babe\c{s}-Bolyai University, Faculty of Mathematics and Computer Science Kog\u{a}lniceanu Street 1, 600410, Cluj-Napoca, Romania,
	Email: pablaga@cs.ubbcluj.ro}
\affil[3]{School of Physics, Sun Yat-Sen University, Xingang Road, Guangzhou 510275, People's Republic of China,
Email: t.harko@ucl.ac.uk}


\begin{document}
\maketitle

\begin{abstract}
The circular restricted three body problem, which considers the dynamics of an infinitesimal particle in the presence of the gravitational interaction with two massive bodies moving on circular orbits about their common center of mass, is a very useful model for investigating the behavior of real astronomical objects in the Solar System. In such a system, there are five Lagrangian equilibrium points, and one important characteristic of the motion is the existence of linearly stable equilibria at the two equilibrium points that form equilateral triangles with the primaries, in the plane of the primaries' orbit. We analyze the stability of motion in the restricted three body problem by using the concept of Jacobi stability, as introduced and developed in the  Kosambi-Cartan-Chern (KCC) theory. The KCC theory is  a differential geometric approach to the variational equations describing the deviation of the whole trajectory of a dynamical system with respect to the nearby ones. We obtain the general result that, from the point of view of the KCC theory and of Jacobi stability, all five Lagrangian equilibrium points of the restricted three body problem are unstable.
\end{abstract}

\section{Introduction}

The three-body problem, the study of the dynamics of three bodies due to their gravitational interactions, is one of the fundamental problems in mathematical physics. One particular case of the general three body problem is the so-called \emph{restricted three body problem}, in which the mass of one of the bodies can be neglected.  Consider two massive bodies that jointly rotate around their common center of mass, with a constant angular velocity $\upomega$. If a third object of negligible mass is placed at rest  with respect to a corotating frame of reference, then when viewed from the noninertial coordinate system it will be initially under the influence of three forces: two due to gravitational interactions with the massive bodies in the system, and a third being the centrifugal force (see \cite{murray1}, \cite{murray2}). The small mass particle does not influence the dynamical motion of the two massive bodies, although they influence its motion. In 1772 Joseph Louis Lagrange has shown that there are five special points in the rotational plane at which the gravitational and the centrifugal forces are in balance. These points are the so-called \emph{Lagrangian equilibrium points} and are designated by $L_1$, $L_2$, $L_3$, $L_4$ and $L_5$, respectively. The first three points are located on a line passing through the massive bodies, while the remaining two are situated on either side of this line and equidistant from the massive bodies, in the orbital plane of the primaries. The behavior of a particle that is given a small deviation from the equilibrium position can be investigated generally by linearising the equations of motion, and carrying out a linear (Lyapounov) stability analysis. The linear stability analysis of the Lagrangian equilibrium points has been intensively investigated, and it has been shown that stable equilibrium occurs only at $L_4$ and $L_5$, for specific values of the mass parameters (see \cite{murray1}, \cite{murray2}, \cite{stab3}). A trapped particle would remain in the neighborhood of these points even it is slightly perturbed. There is some observational evidence that these theoretical concepts have important astronomical implications. In the case of the Sun-Jupiter system there have been discovered small bodies accumulated near  the $L_4$ and $L_5$ points of the system (Trojan asteroids). The two asteroid groups are observed from Earth to lie on either side of Jupiter and to share its synodic period \citep{tr}. It was also speculated that dust may be located in the stable equilibrium points of the Earth-Moon system.

From a purely mathematical point of view, the restricted three body problem is described by a system of two, second order, nonlinear ordinary differential equations. A system of second order differential equations can be investigated by using differential geometric methods, and hence described in geometric terms, by using the general path-space theory of Kosambi-Cartan-Chern (KCC-theory), stimulated and influenced by the geometry of the Finsler spaces (\cite{Ko33}, \cite{Ca33}, \cite{Ch39}, \cite{BM}). The KCC theory describes and interprets the variational equations for the deviation of the whole trajectory with respect to nearby ones in geometric terms. After introducing a non-linear connection, as well as a Berwald type connection, the differential system of equations can be geometrized, and five geometrical invariants can be constructed. The second invariant describes the Jacobi stability properties of the dynamical nonlinear system \citep{Sa05,Sa05a}. The KCC theory has been extensively applied for the study
of various physical, biological, biochemical, or  engineering systems (see \cite{Sa05}, \cite{Sa05a}, \cite{AIM93}, \cite{An03}, \cite{YaNa07} and references therein) and in astronomy and cosmology (see \cite{Har}, \cite{BHS12}, \cite{Ab12}, \cite{Ab13}, \cite{Dan}, and \cite{Lake}, respectively).

It is the goal of this work to investigate the stability of the dynamical equations of motion in the restricted three body problem by using the Kosambi-Cartan-Chern theory, and the associated concept of Jacobi stability. In particular, we consider the Jacobi stability properties of the collinear Lagrangian points  $L_1$, $L_2$, $L_3$, as well as of the triangular Lagrangian points $L_4$ and $L_5$, respectively. As a general result {\it we show that from the perspective of the KCC theory, all the Lagrange points are unstable with respect to the perturbations of the orbits}.

The present paper is organized as follows. We formulate the restricted three-body problem in Section~\ref{sect1}, where we also obtain the critical points of the system, and discuss their linear stability properties. We briefly review the KCC theory and the concept of Jacobi stability  in Section~\ref{kcc}. The Jacobi stability properties of the critical points of the restricted three-body problem are analyzed in Section~\ref{sect2}. We concisely discuss and conclude our results in Section~\ref{sect3}.

\section{The restricted three-body problem}\label{sect1}

Let us consider two celestial bodies with masses $m_{1}$ and $m_{2}$, respectively, with $m_{1}>m_{2}$. Let $\upmu _{1}=m_{1}/\left(m_{1}+m_{2}\right) $ and $\upmu _{2}=m_{2}/\left( m_{1}+m_{2}\right) $ denote the associated reduced masses. A third object with mass $m_{0}<<m_{1}$ has negligible effect on the motions of $m_{1}$ and $m_{2}$, which consequently simply rotates with uniform angular velocity $\upomega $ about their common center of mass. We now establish a coordinate system whose origin coincides with the center of mass, and which spins so that its $x$ axis always passes through $m_{1}$ and $m_{2}$. Let us measure distances in units of the fixed separation of $m_{1}$ and $m_{2}$, and time in units of $\upomega ^{-1}$. In these units, the $x$ coordinate of $\upmu _{1}$ is just $-\upmu _{2}$; conversely, $\upmu _{2}$ is located at $x=\upmu _{1}$. When restricted to the $x-y$ plane, the equations of motion of $m_{0}$ are (see, for example, \cite{murray1})
\begin{equation}\label{eq1}
\ddot{x}-2\dot{y}=x-\left[ \frac{\upmu _{1}\left( x+\upmu _{2}\right) }{r_{1}^{3}%
}+\frac{\upmu _{2}\left( x-\upmu _{1}\right) }{r_{2}^{3}}\right] ,
\end{equation}
\begin{equation}\label{eq2}
\ddot{y}+2\dot{x}=y-\left[ \frac{\upmu _{1}}{r_{1}^{3}}+\frac{\upmu _{2}}{%
	r_{2}^{3}}\right] y ,
\end{equation}
where
\begin{equation}
r_{1}^{2}=\left( x+\upmu _{2}\right) ^{2}+y^{2},
\end{equation}
\begin{equation}
r_{2}^{2}=\left( x-\upmu _{1}\right) ^{2}+y^{2}.
\end{equation}

By introducing the function
\begin{equation}\label{U}
U\left( x,y\right) =\frac{\upmu_1}{r_1}+\frac{\upmu_2}{r_2}+\frac{1}{2}\left(x^2+y^2 \right),
\end{equation}
we can represent Eqs.~(\ref{eq1}) and (\ref{eq2}) as
\begin{equation}\label{eq3}
\ddot{x}-2\dot{y}=\frac{\partial U\left( x,y\right) }{\partial x}%
\end{equation}
\begin{equation}\label{eq4}
\ddot{y}+2\dot{x}=\frac{\partial U\left( x,y\right) }{\partial y}%
\end{equation}

From Eqs.~(\ref{eq3}) and (\ref{eq4}) we obtain the Jacobi first integral (see for example, \cite{murray1})
\begin{equation}
\dot{x}^{2} + \dot{y}^{2} = 2 U - C_J,
\end{equation}
where $C_J$ is a constant of motion.



The position of the equilibrium points is given by the solution of the system of Eqs.~(\ref{eq3}) and (\ref{eq4}) with $\dot{x}=\dot{y}=\ddot{x}=\ddot{y}=0$. There are five different points, denoted by $(x_0, y_0)$, in which these conditions are fulfilled (see for example \cite{br}): the Lagrangian collinear points $L_1$, $L_2$, $L_3$, with $y_0=0$, and $L_{4}$ and $L_{5}$ the triangular Lagrangian equilibrium points, for which $r_{1}=r_{2}=1$. The position of the triangular Lagrangian equilibrium points $L_{4}$ and $L_{5}$ is given by $x_{0}=1/2-\upmu _{2}$ and $y_{0}=\pm \sqrt{3}/2$. For the Lagrangian collinear points y-coordinate is simple $y_0=0$, but the explicit form of the $x-$coordinate of these points is rather lengthly (see, for example, \cite{murray2} for the explicit value for the equilibrium $x-$coordinate for these points and the conditions satisfied by $r_1$, $r_2$ and their partial derivatives with respect to $x$).

\subsection{Linear stability analysis of the equilibrium points of the classical problem}

By introducing the new variables $\left( u,v\right) $ according to the definition $\dot{x}=u$, $\dot{y}=v$, the system of equations Eqs. (\ref{eq1}) - (\ref{eq2}) can be written as a system of first order nonlinear differential
equations as
\begin{equation}
\dot{x}=u \,, \qquad \dot{u}=2v+\frac{\partial U\left( x,y\right) }{\partial x} ,  \label{li1}
\end{equation}
\begin{equation}
\dot{y}=v \,, \qquad \dot{v}=-2u+\frac{\partial U\left( x,y\right) }{\partial y}.  \label{li2}
\end{equation}

The Jacobian matrix of the above system is given by
\begin{equation}
J=\left(
\begin{array}{cccc}
0 & 1 & 0 & 0 \\
\frac{\partial ^{2}U}{\partial x^{2}} &
0 & \frac{\partial ^{2}U}{\partial x\partial y} & 2\\
0 & 0 & 0 & 1 \\
\frac{\partial ^{2}U}{\partial x\partial y}
& -2 & \frac{\partial ^{2}U}{\partial y^{2}} & 0
\end{array}
\right).
\end{equation}
The characteristic equation for the eigenvalues $\uplambda $ of the Jacobian is obtained as
\begin{equation}
{\uplambda }^4 + \left( 4 - {U_{,xx}} - {U_{,yy}} \right) {\uplambda }^2 +
{U_{,xx}}{U_{,yy}}- {{U_{,xy}}}^2 =0,
\end{equation}
where a comma denotes the partial derivative with respect to the given independent variable. Equivalently, the characteristic equation can be written as
\be\label{eigen17}
\uplambda ^4+A\uplambda ^2-B=0,
\ee
where we have denoted $A= \left( 4 - {U_{,xx}} - {U_{,yy}} \right)$ and $B={{U_{,xy}}}^2-{U_{,xx}}{U_{,yy}}$, respectively.

The  solution of the characteristic equation is
\begin{equation}
{{\uplambda ^2}=
	{\frac{{U_{,xx}} +
			{U_{,yy}-4} \pm
			{\sqrt{{\left( {U_{,xx}} +
						{U_{,yy}}-4 \right) }^2 +
					4\left( {{U_{,xy}}}^2 -
					{U_{,xx}}{U_{,yy}}
					\right) }}}{2}}}\,.
\end{equation}

Then for the various derivatives we obtain
\begin{equation}
U_{,xx}(x,y)=1+\frac{\text{$\upmu $}_{1}\left[ 2\left( x+\text{$\upmu $}%
	_{2}\right) ^{2}-y^{2}\right] }{r_{1}^{5}}+\frac{\text{$\upmu $}_{2}\left[
	2\left( x-\text{$\upmu $}_{1}\right) ^{2}-y^{2}\right] }{r_{2}^{5}},
\end{equation}%
\begin{equation}
U_{,yy}(x,y)=1-\frac{\text{$\upmu $}_{1}\left[ \left( x+\text{$\upmu $}%
	_{2}\right) ^{2}-2y^{2}\right] }{r_{1}^{5}}-\frac{\text{$\upmu $}_{2}\left[
	\left( x-\text{$\upmu $}_{1}\right) ^{2}-2y^{2}\right] }{r_{2}^{5}},
\end{equation}%
\begin{equation}
U_{,xy}(x,y)=\frac{3\text{$\upmu $}_{1}y\left( x+\text{$\upmu $}_{2}\right) }{r_{1}^{5}%
}+\frac{3\text{$\upmu $}_{2}y\left( x-\text{$\upmu $}_{1}\right) }{r_{2}^{5}}.
\end{equation}

For the collinear Lagrange points, using the fact that $y=0$, we find
\begin{equation}
U_{,xx}(x,0)=1+\frac{2\text{$\upmu _{1}$}}{\left( x+\text{$\upmu $}_{2}\right)
	^{3}}+\frac{2\text{$\upmu $}_{2}}{\left( x-\text{$\upmu $}_{1}\right) ^{3}},
\end{equation}%
\begin{equation}
U_{,yy}(x,0)=1-\frac{\text{$\upmu $}_{1}}{\left( x+\text{$\upmu $}_{2}\right) ^{3}%
}-\frac{\text{$\upmu $}_{2}}{\left( x-\text{$\upmu $}_{1}\right) ^{3}},
\end{equation}%
\begin{equation}
U_{,xy}(x,0)=0.
\end{equation}

It can be easily shown that $U_{,xx}(x,0) > 0$ and $U_{,yy}(x,0) < 0$ \citep{Meyer}. Hence it follows that the constant term $B$
in Eq.~(\ref{eigen17}) is positive, $B>0$. Since $B > 0$, the solution of Eq.~(\ref{eigen17})  is
\be
\uplambda =-\frac{A}{2}\pm \sqrt{\frac{A^2}{4}+B},
\ee
and hence $J$ has two real and two purely imaginary eigenvalues. Therefore, from
Lyapunov's indirect theorem \citep{Swaters}, it follows that the collinear equilibrium points are unstable saddle points.

Let's consider now the stability of the equilateral Lagrangian points $L_4$ and $L_5$, for which $x_0=1/2-\upmu _2$, and $y_0=\pm \sqrt{3/2}$. Then we obtain first for the eigenvalues of $\uplambda $ the expressions  $\uplambda _{1,2}=\pm \sqrt{-1+27\upmu _2/4}$ and $\uplambda _{3,4}=\sqrt{-27\upmu _2/4}$, respectively. This means that the eigenvalues for the triangular Lagrange points consist of two purely imaginary pairs, and therefore this points are linearly stable with respect to small perturbations. In fact it can be shown that $L_4$ and $L_5$ are linearly stable if the condition $\upmu _1\upmu _2<1/27$ is satisfied \citep{br}.

\section{Brief review of Kosambi-Cartan-Chern (KCC) theory, and of Jacobi stability}\label{kcc}

In this Section we summarize the fundamentals of the KCC-theory extensively used in the following Sections. A complete description of the theory can be found in \cite{AIM93,Sa05, An03}.

Let us consider a real, smooth $n$-dimensional manifold $\mathcal{M}$, and let $T \mathcal{M}$ denote its tangent bundle. On an open connected subset $\Omega $ of the Euclidian $(2n+1)$-dimensional space $R^{n}\times R^{n}\times R^{1}$ we introduce a  $2n+1$ coordinates system $\left( x^{i}\right) =\left( x^{1},x^{2},...,x^{n}\right) $, $\left( y^{i}\right) =\left(
y^{1},y^{2},...,y^{n}\right) $, where by $y^i$ we have denoted
\begin{equation}
\left(y^{i}\right)=\left( \frac{dx^{1}}{dt},\frac{dx^{2}}{dt},...,\frac{dx^{n}}{dt} \right).
\end{equation}
Moreover, by $t$ we denote the time coordinate.

On $\Omega $ we consider now an arbitrary system of second order differential equations, given generally as
\begin{equation}
\frac{\deriv ^{2}x^{i}}{\deriv t^{2}}+2G^{i}\left( x^{j},y^{j},t\right)
=0\,\,,\quad i=\overline{1,n},  \label{EM}
\end{equation}
 where $G^{i}\left( x^{j},y^{j},t\right) $, $i =\overline{ 1,n}$ is a $C^{\infty}$ function in a neighborhood of the initial conditions $\left( x_0,y_0,t_0\right) \in  \Omega$.

One can prove that the system~(\ref{EM}) is generated by a vector field $S$, called semispray,
\begin{equation}
S=y^{i}\frac{\partial }{\partial x^{i}}-2G^{i}\left( x^{j},y^{j},t\right)
\frac{\partial }{\partial y^{i}},
\end{equation}
which introduces a non-linear connection $N_{j}^{i}$, defined as
\begin{equation}
N_{j}^{i}=\frac{\partial G^{i}}{\partial y^{j}}.
\end{equation}
For more details see, for example, \citet{MHSS}.

For a vector field $\upxi ^{i}(x)$  one can define the KCC-covariant differential of $\upxi ^{i}(x)$ on an open subset $\Omega \subseteq R^{n}\times R^{n}\times R^{1}$ according to,
\begin{equation}
\frac{D\upxi ^{i}}{\deriv t}=\frac{\deriv \upxi ^{i}}{\deriv t}+N_{j}^{i}\upxi ^{j}.  \label{KCC}
\end{equation}
For rigorous introductions of the KCC covariant derivative see \cite{AIM93,An03,Sa05,Sa05a}.

Let $x^{i}(t)$ be a trajectory of the system (\ref{EM}), and $\tilde{x}^{i}\left( t\right)$ a small variation into nearby ones, defined by
\begin{equation}
\tilde{x}^{i}\left( t\right) =x^{i}(t)+\upeta\upxi ^{i}(t),  \label{var}
\end{equation}
with $|\upeta|<<1$ a small parameter, and $\upxi ^{i}(t)$ representing the
components of a contravariant vector field, defined along the trajectory $x^{i}(t)$. Replacing~(\ref{var}) into~(\ref{EM}) and taking the limit $\upeta\rightarrow 0$, we get the variational equations describing the perturbations of the orbit (for more details see, for example \cite{AIM93,An03,Sa05,Sa05a}),
\begin{equation}
\frac{\deriv ^{2}\upxi ^{i}}{\deriv t^{2}}+2N_{j}^{i}\frac{\deriv \upxi ^{j}}{\deriv t}+2\frac{\partial
	G^{i}}{\partial x^{j}}\upxi ^{j}=0.  \label{def}
\end{equation}

 Eq.~(\ref{def}) could be written in the covariant form
\begin{equation}
\frac{D^{2}\upxi ^{i}}{\deriv t^{2}}=P_{j}^{i}\upxi ^{j},  \label{JE}
\end{equation}
using the KCC-covariant differential, where
\begin{equation}
P_{j}^{i}=-2\frac{\partial G^{i}}{\partial x^{j}}-2G^{l}G_{jl}^{i}+ y^{l}%
\frac{\partial N_{j}^{i}}{\partial x^{l}}+N_{l}^{i}N_{j}^{l}+\frac{\partial
	N_{j}^{i}}{\partial t},
\end{equation}
and $G_{jl}^{i}\equiv \partial N_{j}^{i}/\partial y^{l}$ is the
Berwald connection (see for example \cite{AIM93}, \cite{MHSS}, \cite{Sa05}, \cite{Sa05a}). We recall that Eq. (\ref{JE}) is the Jacobi equation, and $P_{j}^{i}$ is called the second KCC-invariant, or, alternatively, the deviation curvature tensor.

At last we recall the notion of  Jacobi stability of a dynamical system (see \cite{An03,Sa05,Sa05a}).

\textbf{Definition.} {\it Let the system of ordinary differential equations (\ref{EM}) be given. We assume that with respect to the norm $\left| \left| .\right| \right| $, which is induced by an inner product positive definite, the system
satisfies the initial conditions $\left| \left| x^{i}\left( t_{0}\right) -
\tilde{x}^{i}\left( t_{0}\right) \right| \right| =0$, and $\left| \left| \dot{x}
^{i}\left( t_{0}\right) -\tilde{x}^{i}\left( t_{0}\right) \right| \right|
\neq 0$, respectively. If the real parts of the eigenvalues of the
deviation tensor $P_{j}^{i}$ are strictly negative everywhere, then the trajectories of the system (\ref{EM})
are called \emph{Jacobi stable}. They are called \emph{Jacobi
unstable}, if this condition does not hold.}

For a complete description of the KCC theory see \cite{An03}.

\section{Jacobi stability analysis of the classical restricted three body problem}\label{sect2}

By denoting $x=x^1$, $y=x^2$, $\dot{x}=y^1$, and $\dot{y}=y^2$, Eqs.~(\ref{eq1}) and (\ref{eq2}) giving the equations of motion for the classical restricted three body problem can be written in a form similar to Eqs.~(\ref{EM}) as
\begin{equation}
\frac{\deriv^{2}x^i}{\deriv t ^{2}}+2G^{i}\left( x^1,x^2,y^1,y^2\right) =0, \, \forall i\in\{1,2\},
\end{equation}
where
\begin{equation}
G^{i}\left( x^{1},x^{2},y^{1},y^{2}\right) =\upvarepsilon ^i_jy^{j}-\frac{1}{2}\frac{\partial
	U\left( x^{1},x^{2}\right) }{\partial x^{i}},\, \, \forall i,j \in \{1,2\}\,,
\end{equation}
with the permutation symbol  $\upvarepsilon _{j}^{i}$ defined as $%
\upvarepsilon _{2}^{1}=-\upvarepsilon _{1}^{2}=-1$, and $\upvarepsilon
_{1}^{1}=\upvarepsilon _{2}^{2}=0$.  The nonlinear connection $N_{j}^{i}$ associated to
Eqs.~(\ref{eq1}) and (\ref{eq2}) is given by
\begin{equation}
N_{j}^{i}=\frac{\partial G^{i}\left( x^{1},x^{2},y^{1},y^{2}\right) }{%
	\partial y^{j}}=\upvarepsilon _{j}^{i},\, \forall i,j \in\{1,2\} \,.
\end{equation}
The Berwald connection can be obtained as
\begin{equation}
G_{jl}^{i}=\frac{\partial N_{j}^{i}}{\partial y^{l}},\,\forall i,j,l \in \{1,2\} \,.
\end{equation}
The deviation tensor $P_{j}^{i}$ is given by
\begin{equation}\label{Pij}
P_{j}^{i}=\frac{\partial ^{2}U}{\partial x^{i}\partial x^{j}}+\upvarepsilon
_{l}^{i}\upvarepsilon _{j}^{l}\,.
\end{equation}



Hence it follows that in the case of the classical restricted three body problem the deviation tensor takes the simple form given by  Eq.~(\ref{Pij}). Moreover,  all the other KCC invariants of the system are identically equal to zero. The characteristic equation of the deviation tensor is
\begin{equation}\label{char}
\uplambda^2-(P^1_1+P^2_2) \uplambda + P^1_1 P^2_2 - P^1_2 P^2_1=0	
\end{equation}	
and its eigenvalues are given by
\begin{equation}
\uplambda _{\pm }=\frac{1}{2}\left[ P_{1}^{1}+P_{2}^{2}\pm \sqrt{\left(
	P_{1}^{1}-P_{2}^{2}\right) ^{2}+4P_{1}^{2}P_{2}^{1}}\right] \,.
\end{equation}
The components of the deviation tensor can be explicitly written as
\begin{equation}\label{ctd}
P_{1}^{1}=\frac{\partial ^{2}U}{\partial x^{1}\partial x^{1}}%
-1,P_{2}^{1}=P_{1}^{2}=\frac{\partial ^{2}U}{\partial x^{1}\partial x^{2}},P_{2}^{2}=\frac{\partial ^{2}U}{\partial x^{2}\partial x^{2}}%
-1.
\end{equation}
Thus,  for the eigenvalues of the deviation tensor we find the explicit
expression
\begin{equation}\label{lpm}
\uplambda _{\pm }=\frac{1}{2}\left[ \Delta _{+}U-2\pm \sqrt{\left( \Delta
	_{-}U\right) ^{2}+4\left( \frac{\partial ^{2}U}{\partial x^{1}\partial x^{2}}%
	\right) ^{2}}\right] ,
\end{equation}
where we have denoted
\begin{equation}
\Delta _{+}U=\frac{\partial ^{2}U}{\partial x^{1}\partial x^{1}}+\frac{%
	\partial ^{2}U}{\partial x^{2}\partial x^{2}},\Delta _{-}U=\frac{\partial
	^{2}U}{\partial x^{1}\partial x^{1}}-\frac{\partial ^{2}U}{\partial
	x^{2}\partial x^{2}}.
\end{equation}

By taking into account the explicit form of the potential $U$, we obtain after a simple calculation
\begin{equation}
\Delta _{+}U-2=\frac{\upmu _{1}}{r_{1}^{3}}+\frac{\upmu _{2}}{r_{2}^{3}},
\end{equation}
\begin{equation}
\Delta _{-}U=\frac{3\upmu _{1}\left( x^{1}-x^{2}+\upmu _{2}\right) \left(
	x^{1}+x^{2}+\upmu _{2}\right) }{r_{1}^{5}}+\frac{3\upmu _{2}\left(
	x^{1}+x^{2}-\upmu _{1}\right) \left( x^{1}-x^{2}-\upmu _{1}\right) }{r_{2}^{5}},
\end{equation}
and
\begin{equation}
\frac{\partial ^{2}U}{\partial x^{1}\partial x^{2}}=3x^{2}\left[ \frac{\upmu
	_{1}\left( x^{1}+\upmu _{2}\right) }{r_{1}^{5}}+\frac{\upmu _{2}\left( x^{1}-\upmu
	_{1}\right) }{r_{2}^{5}}\right] ,
\end{equation}
respectively. In the case of the $L_4$ and $L_5$ Lagrangian points we have $r_1=r_2=1$ and $x^1=1/2-\upmu _2$ and $x^2=\pm \sqrt{3}/2$, respectively. Thus, we immediately obtain the eigenvalues of the deviation curvature tensor at these points as
\begin{equation}\label{L45}
\uplambda _{\pm }^{L_{4},L_{5}}=\frac{1}{2}\pm \frac{3}{4}\sqrt{1+3\left( 1-2\upmu
	_{2}\right)^{2}}.
\end{equation}
In the case of the collinear Lagrangian points, since $\partial ^{2}U/\partial x^{1}\partial x^{2}|_{x^2=0}\equiv 0$, we obtain
\begin{equation}\label{L123}
\uplambda _{+}^{L_{1},L_{2},L_{3}}=\frac{1}{2}\left.\left( \Delta _{+}U-2\pm \Delta _{-}U\right)\right|_{x^2=0},
\end{equation}
and thus
\begin{equation}
\uplambda _{+}^{L_{1},L_{2},L_{3}}=\left. \left( \frac{\partial ^{2}U}{%
	\partial x^{1}\partial x^{1}}-1\right) \right| _{x^2=0}=2\left[ \frac{\upmu _{1}%
}{\left( x^1+\upmu _{2}\right) ^{3}}+\frac{\upmu _{2}}{\left( x^1-\upmu _{1}\right)
	^{3}}\right] ,
\end{equation}
\begin{equation}
\uplambda _{-}^{L_{1},L_{2},L_{3}}=\left. \left( \frac{\partial ^{2}U}{%
	\partial x^{2}\partial x^{2}}-1\right) \right| _{x^2=0}=- \frac{\upmu _{1}%
}{\left( x^1+\upmu _{2}\right) ^{3}}-\frac{\upmu _{2}}{\left( x^1-\upmu _{1}\right)
	^{3}} .
\end{equation}

To establish the sign of the real part of the eigenvalues of characteristic equation~({\ref{char}}), we use the sign of its discriminant and Vi\`ete's formulas. The discriminant of~({\ref{char}}) is the term under square root in~(\ref{lpm}). Taking into account the explicit form of the potential $U$ from~(\ref{U}), after straightforward calculations, we get
\begin{equation}\label{D}
\Delta = \frac{9\upmu_1^2}{r_1^6} + \frac{9 \upmu_2^2}{r_2^6} + \frac{18 \upmu_1 \upmu_2}{r_1^5 r_2^5} \{[(x^1+\upmu_2)(x^1-\upmu_1)+(x^2)^2]^2-(x^2)^2\}\,.
\end{equation}

Using Vi\`ete's formulas, we determine the sum and the product of the roots of the characteristic equation~({\ref{char}}). The sum, $S$, is
\begin{equation}
S=P^1_1+P^2_2
\end{equation}
and the product, $P$, is
\begin{equation}
P=P^1_1 P^2_2 - P^1_2 P^2_1\,.
\end{equation}

Taking into account the form of the components of the deviation tensor~(\ref{ctd}) and the expression of the potential $U$, given by~(\ref{U}), the sum and the product of the roots of the characteristic equation~({\ref{char}}) are
\begin{equation}\label{SP}
S=\frac{\upmu_1}{r_1^3}+\frac{\upmu_2}{r_2^3}\,,\quad P=-2\left( \frac{\upmu_1}{r_1^3}+\frac{\upmu_2}{r_2^3} \right)^2 + \frac{9 \upmu_1 \upmu_2}{r_1^5 r_2^5} (x^2)^2\,.
\end{equation}

In the $L_4$ and $L_5$ Lagrangian points, $x^1=\frac{1}{2}-\upmu_2$, $x^2=\pm \frac{\sqrt{3}}{2}$ and $r_1=r_2=1$. Replacing these values in~(\ref{D}) and~(\ref{SP}) we get
\begin{equation}
\Delta = 9 (3 \upmu_2^2 - 3 \upmu_2 +1)\,,\quad P= \frac{1}{4} (-27 \upmu_2^2 + 27 \upmu_2 - 8 )\,,\quad S=1\,.
\end{equation}

It is easy to prove that $\Delta$ is always positive and $P$ is negative, for all real values of $\upmu_2$. Therefore, the eigenvalues of the characteristic equation~(\ref{char}), in $L_4$ and $L_5$ Lagrangian points~(\ref{L45}), are real numbers, with opposite signs. Based on the theory developed in Section~\ref{kcc}, {\it we conclude that the triangular Lagrangian equilibrium points of the restricted three body problem are Jacobi unstable}.

In the case of the collinear Lagrangian points, $L_1$, $L_2$, $L_3$, the discriminant of the characteristic equation~(\ref{char}) reduces to a nonzero perfect square
\begin{equation}
\Delta = 9 \left( \frac{\upmu_1}{r_1^3} + \frac{\upmu_2}{r_2^3}\right)^2 \,,
\end{equation}
meaning that the roots of the characteristic equation are always real numbers. The product of the eigenvalues is
\begin{equation}
P = - 2 \left( \frac{\upmu_1}{r_1^3} + \frac{\upmu_2}{r_2^3}\right)^2 \,,
\end{equation}
a negative, nonzero, real number. Therefore, we reach the conclusion that {\it the collinear Lagrangian equilibrium points in the restricted three body problem are Jacobi unstable}.

\section{Discussions and final remarks}\label{sect3}

As we have mentioned in introduction, the linear analysis stability of the Lagrangian equilibrium points was intensively investigated. The studies demonstrated that the collinear points $L_1$, $L_2$, $L_3$ are unstable for all $\upmu_2<1/2$, but in special initial conditions, the third body  could describe a stable, periodic orbit \citep{Sz67}. For example, the spacecraft of SOHO mission describes a periodic orbit around the Lagrangian point $L_1$ of the Earth-Sun system \citep{Dom95}. {\it The Jacobi stability analysis revealed that the collinear Lagrangian equilibrium points of the restricted three body problem are unstable for all values of $\upmu_2$}.

In the case of the triangular Lagrangian equilibrium points $L_4$ and $L_5$, the stability was proven for $\upmu_2 \leq \frac{27-\sqrt{621}}{54} \approx 0.0385$ \citep{br}, excepting few specific values, where there is a possibility for resonances to appear \citep{ddb}. Performing the Jacobi stability analysis we obtained that {\it the triangular Lagrangian equilibrium points of the restricted three body problem  are unstable for all values of $\upmu_2$}.

The concepts and methods of the KCC theory are very well suited for the description of the geometric and stability properties of the dynamical systems. There is a fundamental difference between the classic linear stability analysis, and the Jacobi one. The Lyapunov, or linear, stability analysis is based on the linearization near the critical points of a dynamical system via the Jacobian matrix of the nonlinear system. On the other hand, the KCC theory considers the stability of an entire bunch of trajectories in a tubular region around the considered path \citep{Sa05}.

The Jacobi analysis revealed that the Lagrangian equilibrium points are unstable for all values of the mass parameter $\upmu_2$. This instability can be interpreted in the sense that the trajectories of the particles in the restricted three body problem in the given coordinate system will scatter when approaching the initial point. The Jacobi stability of the trajectories of the restricted three body problem can also be regarded as characterizeing the {\it robustness of the system} with respect to the small perturbations of the entire trajectory. We can also regard the Jacobi stability as describing the {\it resistance of the entire trajectory} to the emergence of the chaotic dynamics due to small perturbations of the system. In this context it is worth to point out again that {\it all Lagrangian points of the restricted three body problem are Jacobi unstable}.

It is also interesting to note that our results show a complex relation between the linear (Lyapounov) stability analysis of the Lagrangian critical points of the restricted three body problem, and the robustness of the whole system trajectories with respect to small perturbations, as described by the Jacobi stability via the deviation curvature tensor. If the Lyapounov stability analysis introduces regions of stability/instability, determined by the numerical values of the parameter $\upmu _2$, the Jacobi instability is independent on the numerical values of $\upmu _1$ and $\upmu _2$.

In the present paper we have performed a full stability analysis of the circular restricted three body problem, in which we have considered a description of the deviations of the full bunch of trajectories of the complex nonlinear differential system describing the motion of the three gravitationally interacting particles.  Moreover, we have developed some basic theoretical and computational tools for the in depth study of this type of stability, and its applications to astronomical problems. Further investigations of the Jacobi stability properties of the three body and other celestial mechanical problems may provide some powerful analytical methods for describing the stability and evolutionary properties of the gravitationally interacting many body systems, and help the better understanding of the dynamical characteristics of the  particle motion.



\begin{thebibliography}{}

	\bibitem[Abolghasem (2012)]{Ab12}
	Abolghasem H.: 2012, \textit{Journal of Dynamical Systems and Geometric Theories}, \textbf{10}, 197.
	
	\bibitem[Abolghasem (2013)]{Ab13}
	Abolghasem H.: 2013, \textit{International Journal of Differential Equations and Applications}, \textbf{12}, 131.
	
	\bibitem[Abhyankar (1959)]{stab4}
	 Abhyankar K.D.: 1959, \emph{Astron. J.} \textbf{64}, 163.

	\bibitem[Antonelli and Bucataru (2003)]{An03}
	Antonelli P.L. and Bucataru I.: 2003 \textit{in} \emph{Handbook of Finsler geometry}, vol. \textbf{1}, 83, Kluwer Academic, Dordrecht.
	
	\bibitem[Antonelli \etal (1993)]{AIM93}
	Antonelli, P. L., Ingarden, R. S., Matsumoto, M.: 1993, \textit{The Theory of Sprays and Finsler Spaces with Applications in Physics and Biology}, Vol. \textbf{58}, Springer Science+Business Media, Dordrecht.
	
	\bibitem[Boehmer \etal (2012)]{BHS12}
	Boehmer, C.G., Harko, T., Sabau, S.V.: 2012, \textit{Advances in Theoretical and Mathematical Physics}, \textbf{16}, 1145.
	
	\bibitem[Brouwer and Clemence (1961)]{br}
	Brouwer D., Clemence G.M.: 1961, \emph{Methods of Celestial Mechanics}, Academic Press, New York.
	
	\bibitem[Bucataru and Miron (2007)]{BM}
	Bucataru, I., Miron, R.: 2007, \emph{Finsler-Lagrange Geometry: Applications to Dynamical Systems}, Publishing House of the Romanian Academy, Bucharest.
	
	\bibitem[Cartan (1933)]{Ca33}
	Cartan E: 1933, \emph{Math. Z.}, \textbf{37}, 619.

	\bibitem[Chern (1939)]{Ch39}
	Chern S.S.: 1939, \emph{Bulletin des Sciences Mathematiques}, \textbf{63}, 206.

\bibitem[Danila \etal (2016)]{Dan} D\u{a}nil\u{a}, B.,  Harko, T., Mak, M. K., Pantaragphong, P. and Sabau, S.: 2016, \emph{Advances in High Energy Physics} {\bf 2016},  7521464.
	
	\bibitem[Deprit and Deprit-Bartholome (1972)]{ddb}
	Deprit A. and Deprit-Bartholome A.: 1972, \emph{Astron. J}, \textbf{72}, 173.
	
	\bibitem[Domingo, Fleck and Poland (1995)]{Dom95}
	Domingo V., Fleck B. and Poland A.I.: 1995, \emph{Solar Physics}, \textbf{162}, 1.
	
	\bibitem[Duboshin (1938)]{stab3}
	Duboshin G.N.: 1938, \emph{Sov. Astron.}, \textbf{15}, 209.

\bibitem[Harko and Sabau (2008)]{Har} Harko, T. and Sabau, S.: 2008, \emph{Phys. Rev. D}, {\bf  77},  104009.
	
	\bibitem[Kosambi (1933)]{Ko33}
	Kosambi D.D.: 1933 \emph{Math. Z.}, \textbf{37}, 608.

\bibitem[Lake and Harko (2016)]{Lake} Lake, M. J. and Harko, T.: 2016, \emph{The European Physical Journal C}, {\bf 76}, 311.
	
	\bibitem[McCusky (1963)]{stab5}
	McCusky S.W.: 1963, \emph{Introduction to Celestial Mechanics}, Addison-Wesley, New York.

\bibitem[Meyer and Hall (1992)]{Meyer}  Meyer K. R. and Hall G. R.: 1992,  \emph{Introduction to Hamiltonian Dynamical Systems and the
	N-Body Problem},  Springer-Verlag, Berlin, Heidelberg

	\bibitem[Miron and Frigioiu (2005)]{MiFr05}
	Miron R., Frigioiu C.: 2005, \emph{Algebras Groups Geom.}, \textbf{22}, 151.

	\bibitem[Miron \etal (2001)]{MHSS}
	Miron R., Hrimiuc D., Shimada H, Sabau V.S.: 2001, \textit{The Geometry of Hamilton and Lagrange Spaces}, Kluwer Acad. Publ., Dordrecht, Boston.

	\bibitem[Murray (1994)]{murray1}
	Murray C.D.: 1994, \emph{Icarus}, \textbf{112}, 465.

	\bibitem[Murray and Dermott (1999)]{murray2}
	Murray, S.F. Dermott: 1999, \emph{Solar System Dynamics}, Cambridge University Press, Cambridge.

	\bibitem[Sabau (2005a)]{Sa05}
	Sabau S.V.: 2005a, \emph{Nonlinear Analysis}, \textbf{63}, 143.

	\bibitem[Sabau (2005b)]{Sa05a}
	Sabau S.V.: 2005b, \emph{Nonlinear Analysis: Real World Applications}, \textbf{6}, 563.
	
	\bibitem[Swaters (2000)]{Swaters} Swaters G. E.: 2000, \emph{Introduction to Hamiltonian fluid dynamics and stability theory}, Chapman \& Hall CRC, Taylor \& Francis Group, Boca Raton, London, New York
	
	\bibitem[Szebehely (1967)]{Sz67}
	Szebehely V.: 1967, \emph{Theory of Orbits}, Academic Press, New-York and London.
	
	\bibitem[Yajima and Nagahama (2007)]{YaNa07}
	Yajima T., Nagahama H.: 2007, \emph{J. Phys. A: Math.Theor.}, \textbf{40}, 2755.

	\bibitem[Van Houten, Van Houten-Groenewald and Gehrels (1970)]{tr}
	Van Houten C.J., Van Houten-Groenewald I. and Gehrels T.: 1970, \emph{Astron. J.}, \textbf{75}, 659.
		
\end{thebibliography}
\makeatletter
\def\@biblabel#1{}
\makeatother

\received{\it *}
\end{document}